\documentstyle[aasms4]{report}

\newcommand{\be}{\begin{equation}}
\newcommand{\ee}{\end{equation}}

\newcommand{\sfrac}[2]{{\textstyle\frac{#1}{#2}}}

\newcommand{\ba}{\begin{eqnarray}}
\newcommand{\ea}{\end{eqnarray}}
\newcommand{\del}{\nabla}

\newcommand{\etal}{\emph{et~al}.}

\newcommand{\udot}{\dot{u}}
\newcommand{\sdel}{\widetilde{\del}}
\newcommand{\li}{&\!\!\!\!}

\newtheorem{thm}{Theorem}
\newtheorem{cor}{Corollary}

\received{\today}
\accepted{}
\journalid{}{}
\articleid{}{}

\lefthead{C. A. Clarkson and R. K. Barrett}
\righthead{Generalised EGS Theorems}

\begin{document}
\bibliographystyle{astron}
\baselineskip=0.7\baselineskip

\title{Does the Isotropy of the CMB Imply a Homogeneous Universe?\protect\\
       Some Generalised EGS Theorems.}

\author{Chris A. Clarkson\footnote{Email: \tt{chris@astro.gla.ac.uk}}}
\author{Richard K. Barrett\footnote{Email: \tt{richard@astro.gla.ac.uk}}}
\vskip 0.5cm

\affil{Dept. Physics and Astronomy,\\University of Glasgow,
\\University Avenue,\\Glasgow.\\G12 8QQ.\\}

\affil{}
\date{\today}

\begin{abstract}

We demonstrate that the high isotropy of the Cosmic Microwave Background (CMB),
combined with the Copernican principle, is not sufficient to prove homogeneity
of the universe~-- in contrast to previous results on this subject. The crucial
additional factor not included in earlier work is the acceleration of the
fundamental observers. We find the complete class of irrotational perfect fluid
spacetimes admitting an exactly isotropic radiation field for every fundamental
observer and show that they are Friedman-Lema\^\i tre-Robertson-Walker (FLRW)
if and only if the acceleration is zero. While inhomogeneous in general, these
spacetimes all possess three-dimensional symmetry groups, from which it follows
that they also admit a thermodynamic interpretation.

In addition to perfect fluid models we also consider multi-component fluids
containing non-interacting radiation, dust and a quintessential scalar field or
cosmological constant in which the radiation is isotropic for the geodesic
(dust) observers. It is shown that the non-acceleration of the fundamental
observers forces these spacetimes to be FLRW.

While it is plausible that fundamental observers (galaxies) in the real
universe follow geodesics, it is strictly necessary to determine this from
local observations for the cosmological principle to be more than an
assumption. We discuss how observations may be used to test this.

\end{abstract}


\section{Introduction}

The high isotropy of the CMB is usually taken as strong evidence that the
universe is homogeneous and isotropic, i.e.,~is well described by an FLRW
model. The principle justification for this is an important theorem of Ehlers,
Geren and Sachs~(1968)\nocite{ehlers-68} (based on earlier work by Tauber and
Weinberg~1961\nocite{tau-wei61}), which states that if all observers in an
expanding, dust universe measure an isotropic CMB then the universe is FLRW and
the cosmological principle is valid. The importance of this theorem lies in the
fact that it permits the homogeneity and isotropy of the universe to be deduced
not from measurements of the actual isotropy of the universe about us, but from
only measurements of the CMB, combined with the Copernican principle (that is,
the assumption that all observers in the universe see the same degree of
isotropy). The Copernican principle is often regarded as a powerful but
untestable assumption in cosmology, although there are suggestions that
it may be testable using the Sunyaev-Zeldovich effect, for example
(Goodman~1995\nocite{goodman95}). Here we simply assume that the Copernican
principle is valid and study the consequences of applying it to the observed
high degree of isotropy of the CMB. That is, we examine spacetimes with an
isotropic CMB for \emph{all} observers.

The EGS theorem has been generalised by Treciokas and
Ellis~(1971)\nocite{trec-ell71} to include an isotropic collision term.
Ferrando, Morales, and Portilla~(1992)\nocite{ferran-92} find the general form
of the energy-momentum tensor and Einstein's equations for spacetimes with an
isotropic radiation field, and consider some special cases with anisotropic
pressure. It has also been shown by Stoeger, Maartens and
Ellis~(1995)\nocite{sto-maa95} that the EGS theorem almost holds when applied
to an almost isotropic radiation field.

There are counterexamples to the spirit of the EGS theorem (that is, when some
of the assumptions are relaxed the result fails to hold). In particular, Ellis,
Maartens and Nel~(1978)\nocite{ellis-78} show that the result does not hold if
the expansion is zero (which is obviously not relevant to cosmology), and
Ferrando \etal~(1992) emphasise that homogeneity does not follow if there is
anisotropic pressure in the energy-momentum tensor. Most importantly for the
work presented here, though, is the result of
Barrett and Clarkson~(1999a)\nocite{bar-clII99},
which shows that when the assumption of geodesic observers is relaxed there
exist inhomogeneous perfect fluid (or scalar field) cosmologies with an
isotropic CMB. (In fact, we show in this paper that the Stephani models
considered in Barrett and Clarkson~(1999a) are representatives of the only
perfect fluid spacetimes that admit an isotropic CMB for all fundamental
observers.) Nilsson \etal~(1999)\nocite{nils-99} provide a counterexample to
the almost EGS result when the Weyl curvature is not negligible.

The basis of the EGS theorem is the Liouville equation for photons, which tells
us that if a radiation field (i.e.,~a solution of the Liouville equation)
exists such that for every observer on some timelike congruence the radiation
field is isotropic, then that congruence is (parallel to) a conformal Killing
vector (CKV). This may be expressed more formally as follows (Ehlers
\etal~1968; Ferrando \etal~1992):
\begin{thm}
\label{isoradthm}
A spacetime will admit an isotropic radiation field if and only if it is
conformal to a stationary spacetime, which happens if and only if there is a
velocity field~$u^a$ satisfying
\ba
\sigma_{ab}=0,\label{shear=0}\\
\del_{[a}\left(\udot_{b]}-\sfrac{1}{3}\theta u_{b]}\right) = 0,
\label{Qcondition}
\ea
where $\sigma_{ab}$, $\udot^a$ and~$\theta$ are the shear, acceleration and
expansion of~$u^a$, respectively.
\end{thm}
(Then~$u^a$ is the velocity field relative to which the radiation is isotropic,
and is parallel to the CKV.)

In the absence of some statement about the matter content of a spacetime, or
further assumptions about the congruence~$u^a$, Theorem~\ref{isoradthm} is all
that can be said. In a cosmological context the simplest, and most common,
assumption is that the matter is dust (implying that~$u^a$ is geodesic), which
leads to:
\begin{thm}[Ehlers, Geren, and Sachs]
\label{EGSthm}
If the fundamental observers in a dust spacetime see an isotropic radiation
field, then the spacetime is locally FLRW.
\end{thm}

Alternatively, we can simply assume that~$u^a$ is geodesic. The existence of an
isotropic radiation field then ensures (for non-zero expansion) that the energy
flux relative to~$u^a$ is zero. If the anisotropic stress tensor is zero at any
instant (so that the energy-momentum tensor has perfect fluid form) then it
will remain zero and the spacetime will be FLRW (Ferrando \etal~1992, Corollary
1; but note that their statement that the anisotropic stress is invariant along
$u^a$ in general is misleading~-- from Eqs. (31) and (40) of Ellis~1998 we have
$\dot\pi_{\langle ab\rangle}\propto\theta\pi_{ab}$).

It is worth emphasising that in applications of the above results to cosmology
the motion of the fundamental observers must be identified with the
congruence~$u^a$. For example, in~\S\ref{CFsolns} \emph{all} Stephani models
are conformally flat, and therefore conformally stationary, but for most of
these spacetimes the fluid congruence is \emph{not} aligned with the timelike
CKV.

The matter content of the universe is not precisely known. Certainly, there is
a large number of possible contributors, including hot and cold dark matter (in
their various manifestations), electromagnetic fields~etc., as well as the more
obvious radiation and baryonic matter. In particular, the type~Ia supernova
results of Perlmutter~\etal\ (1999)\nocite{perl-99} suggest that an important
component may be a `quintessential' scalar field. However, the forms of matter
that are thought to contribute significantly to the energy-momentum tensor may
be treated in general as perfect fluids. That is, their energy-momentum tensor
may be written in the form
\be
                      T_{ab}=\mu u_au_b+ph_{ab},
\label{perfectfluid}
\ee
where $\mu$ and~$p$ are the energy density and pressure, $u^a$~is the timelike
velocity congruence of the fluid, and $h_{ab}$ is the spatial projection tensor
associated with~$u^a$. Scalar fields may also be written in this form,
with~$u^a$ parallel to the gradient of the scalar field, provided that this is
timelike (see~\S\ref{QCDM}). Note that if several such components are present
there is no reason why their fundamental congruences (the~$u^a$'s) should be
parallel. If they are, then they may be treated as effectively a single perfect
fluid with the energy densities and pressures added together. If not, the
decomposition of the energy-momentum with respect to~$u^a$ for one such fluid
(as in equation~(14) of Ellis~1998) will contain energy flux and anisotropic
stress terms from the other fluids (again, see~\S\ref{QCDM}). The fundamental
observers will be associated with one such congruence. Usually the fundamental
observers in standard models of the universe are associated with a dust-like
($p=0$) component, with the result that the acceleration~$\dot{u}^a$ of the
fundamental congruence is zero. However, we wish to study the consequences of
relaxing this assumption and consider models with acceleration. This
acceleration must be caused by some non-gravitational force (typically pressure
gradients for perfect fluid spacetimes, but in principle it could be the result
of a coupling between the fluid and some other component such as the
electromagnetic field).

With this in mind, in this paper we consider perhaps the two simplest
generalisations of the dust hypothesis. Firstly we imagine that the dominant
form of matter is a single irrotational perfect fluid. We do not specify what
form of matter this corresponds to, but we allow pressure gradients that give
rise to acceleration. Secondly we consider cosmological models in which more
than one matter component makes a significant contribution to the energy
density and dynamics of the universe. Specifically we consider `QCDM' models,
containing a non-interacting mixture of radiation, dust (CDM) and a scalar
field (or a cosmological constant). The observers are associated with the CDM
component, and are therefore geodesic. The difference between this and other
theorems assuming geodesic observers is that the the scalar field component can
introduce effective energy flux and anisotropic stresses relative to the dust
congruence, and so the matter need not behave as a perfect fluid.

In the following section we find all irrotational perfect fluid solutions
admitting an isotropic radiation field for the fundamental observers, showing
that they form a subclass of the Stephani spacetimes and are FLRW if and only
if the acceleration vanishes. Then in~\S\ref{QCDM} we examine QCDM models and
prove that such models must be homogeneous and isotropic if they admit an
isotropic radiation field. Finally, in~\S\ref{concs} we emphasise the
importance of acceleration for these results and show that the acceleration of
the fundamental congruence is, in principle, detectable, and measureable, in
galaxy surveys. Two appendices contain results relating to~\S\ref{CFsolns}.

\section{The Irrotational Perfect Fluid Solutions.}
\label{IPFSolns}

We wish to consider the constraints imposed by the existence of an isotropic
radiation field for the fundamental observers on perfect fluid spacetimes in
which the rotation of the fundamental congruence is zero. Since it follows from
Theorem~\ref{isoradthm} that the shear of the fundamental congruence must also
be zero we immediately know that all the acceptable solutions are members of
the Stephani-Barnes family, which is the family of \emph{all} shear-free,
irrotational, expanding (or contracting) perfect fluids (see Barnes~1973;
Krasi\'{n}ski~1989,~1997\nocite{kras97}\nocite{kras89}). It only remains, then,
to impose condition~(\ref{Qcondition}) of Theorem~\ref{isoradthm} and thus find
the sub-class of Stephani-Barnes models which admit an isotropic radiation
field for all fundamental observers.

The Stephani-Barnes family contains the Barnes solutions, which are of Petrov
type~D, and the Stephani models, which are conformally flat, although these two
classes overlap where the Barnes solutions degenerate to type O (these
solutions then become Stephani models with symmetry). The FLRW models are a
subcase of these solutions. The Barnes spacetimes all possess symmetry (see
below), whereas the Stephani spacetimes, in general, do not. In all cases the
metric in comoving coordinates can be written in the form (see
Krasi\'{n}ski~1989; although here we use the same notation as
Krasi\'{n}ski~1997):
\be
ds^2=V^{-2}\left\{-(FV_{,t})^2dt^2+dx^2+dy^2+dz^2\right\}\label{metric}
\ee
where $F=F(t)$ and $V=V(t,x,y,z)$. $F(t)$ is arbitrary, but there are some
restrictions on the form of $V$ depending on the symmetries of the solution,
and these will be discussed in due course (but the impatient reader may wish to
note Eqs.~\ref{SS-plane-V},~\ref{hyperbolic-V},~\ref{hyp-V-de},
and~\ref{V_stephani}). Expressions for the energy density and pressure can be
found in Krasi\'{n}ski (1997).

The fluid velocity is given by (without loss of generality we can assume
that~$V>0$)
\be
          u^a=\frac{V}{|FV_{,t}|}\delta^a_{\phantom{a}0},
\label{velocity}
\ee
with expansion
\be
\theta=-\hbox{sign}(V_{,t})\frac{3}{|F|},
\label{expansion}
\ee
and acceleration
\be
          \udot_0=0, \hskip 1.0cm
          \udot_i=\frac{\partial}{\partial x^i}\ln\frac{V_{,t}}{V},
\label{acceln}
\ee
where $i=1,2,3$, $x^i=\{x,y,z\}$. Note that~(\ref{expansion}) differs from the
expression usually given for the expansion (in Krasi\'{n}ski 1997, equations
(4.1.4) and~(4.9.6), for example) by the inclusion of the
$-\hbox{sign}(V_{,t})$ factor. Neglect of this factor is inconsistent since~$F$
enters the metric only quadratically, so the sign of the expansion cannot
depend on the sign of~$F$. The sign of~$\theta$ \emph{does} depend on that
of~$V_{,t}$, though: for the Friedmann subcase of the Stephani-Barnes models,
for example, $V$ is related to the scale factor~$R(t)$ by~$V=1/R$,
and~$|FV_{,t}/V|=1$ (see~\S\ref{CFsolns}), so
that~$\theta=3\dot{R}/R=-3V_{,t}/V=-3\,\hbox{sign}(V_{,t})/|F|$. This is
important here because the constraint~(\ref{Qcondition}) contains the
expansion. Note, too, that~$F$ is not a true degree of freedom parameterising
distinct spacetimes, but rather represents a coordinate freedom, corresponding
to different choices of the time coordinate.

From (\ref{velocity}), (\ref{expansion}) and~(\ref{acceln}), the
condition~(\ref{Qcondition}) leads to the constraint:
\be
\frac{\partial^2}{\partial x^i\partial t}\ln V_{,t}=0,
\label{deVcond}
\ee
which is satisfied if and only if the function~$V$ has the form
\be
V(t,x,y,z)=T(t)X(x,y,z)+Y(x,y,z),\label{Vcondition}
\ee
where $T$, $X$, and~$Y$ are arbitrary functions. This equation is the key
additional constraint on the Stephani-Barnes solutions.

It is worth noting that it follows from~(\ref{deVcond}) that the
acceleration scalar is constant along the fluid flow for every observer,
i.e.,~$\udot_{,t}=0$ (where $\udot^2=\udot_a\udot^a$), as can be seen by
calculating~$\udot^2_{,t}$ from~(\ref{acceln}). In fact, it can be verified
more generally that for any conformally stationary spacetime (i.e.,~a
spacetime satisfying (\ref{shear=0}) and~(\ref{Qcondition})), even with
rotation, the acceleration scalar evolves according to
\[
            u^a\del_a \udot^2 = \sfrac23 \udot^b\sdel_b\theta,
\]
where $\sdel$ denotes the spatially projected gradient (see Ellis~1998). It
follows from equation~(32) of Ellis~(1998) that~$\sdel_b\theta=0$ whenever
the rotation vanishes (for a perfect fluid), which is the case for the
Stephani-Barnes models.

Note that if~$V$ has the form~(\ref{Vcondition}) we can immediately write the
metric in a manifestly conformally static form
\[
ds^2=\frac{X^2}{V^2}\left\{-d\tau^2+X^{-2}(dx^2+dy^2+dz^2)\right\},
\]
where $d\tau=T_{,t}F \,dt$, which shows that these models will indeed be
conformally stationary, as required by Theorem~\ref{isoradthm}. We now discuss
each subcase in turn.

\subsection{The Barnes Solutions.}

The Barnes solutions (Barnes~1973) all have spherical, plane or hyperbolic
symmetry (i.e.,~they possess three-dimensional isometry groups acting on
two-dimensional orbits,~\emph{cf}.~\S\ref{CFsolns}). The restrictions on the
metric function~$V$ depend on which of these symmetries the spacetime possesses
(see Krasi\'nski~1997). For the solutions with spherical symmetry (the
Kustaanheimo-Qvist solutions) or planar symmetry we introduce a new independent
variable~$u(x,y,z)$ defined by $u=r^2=x^2+y^2+z^2$ in the spherical case
and~$u=z$ in the planar case. Then~$V$ is defined by~$V=V(t,u)$, subject to the
condition:
\be
\frac{\partial^2 V}{\partial u^2}=f(u)V^2.\label{SS-plane-V}
\ee
where~$f$ is an arbitrary function. Since~$V=V(t,u)$ we know
from~(\ref{Vcondition}) that in order to admit an isotropic radiation field for
all observers~$V$ must have the form
\be
V(t,u)=T(t)X(u)+Y(u).\label{Vcondition2}
\ee
Equation~(\ref{SS-plane-V}) then imposes a constraint on the functions $X$
and~$Y$, which will be outlined below.

For the solutions with hyperbolic symmetry the constraint on~$V$ is very
similar. This time we introduce the variable~$u=x/y$. $V$~can then be written
\be
V(t,x,y)=yW(t,u),
\label{hyperbolic-V}
\ee
where~$W$ satisfies
\be
\frac{\partial^2 W}{\partial u^2}=f(u)W^2,
\label{hyp-V-de}
\ee
with~$f$ once again a free function. The condition~(\ref{Vcondition}) now gives
\[
V(t,x,y)=T(t)X(x,y)+Y(x,y)=yW(t,u).
\]
Dividing by~$y$ and redefining $X$ and~$Y$ in an obvious fashion we obtain
\be
W(t,u)=T(t)\tilde{X}(u)+\tilde{Y}(u),\label{Wcondition}
\ee
in which $\tilde{X}$ and~$\tilde{Y}$ are again constrained by~(\ref{hyp-V-de}).

For all three symmetries of the Barnes solutions, then, the constraints on the
metric function~$V$ essentially reduce to the second-order differential
equations~(\ref{SS-plane-V}) or~(\ref{hyp-V-de}). Imposing the
condition~(\ref{Vcondition}) introduces the additional constraint
(\ref{Vcondition2}) or~(\ref{Wcondition}). Substitution of (\ref{Vcondition2})
or~(\ref{Wcondition}) into (\ref{SS-plane-V}) or~(\ref{hyp-V-de}) respectively
and differentiating twice with respect to time, dividing by~$T_{,t}$ each time
(and recognising that $T_{,t}\ne 0$, $X\ne 0$, so that $V_{,t}\ne 0$
in~(\ref{metric})), leads directly to the condition
\[
                            f(u)=0.
\]
Barnes solutions with~$f(u)=0$ are conformally flat (Krasi\'{n}ski~1997,~p.142)
and are therefore actually a subcase of the Stephani models. That is, proper
Barnes spacetimes can be ruled out: they do not admit an isotropic radiation
field. It only remains to apply the condition~(\ref{Vcondition}) to the
Stephani models, which we do in the next section.

\subsection{The Conformally Flat Solutions}
\label{CFsolns}

The conformally flat sub-case is the entire class of conformally flat,
expanding, perfect fluid solutions, and is the Stephani solution (Stephani
1967a,b).\nocite{step67a,step67b} The function~$V$ is most often written in the
form (see Krasi\'{n}ski~1997;
Barrett and Clarkson~1999a,b\nocite{bar-clI99,bar-clII99}):
\be
                                  V(t,x,y,z) =
   \frac{1}{R(t)}\left(1+\sfrac14 k(t)|{\mathbf{x}}-{\mathbf{x}}_0(t)|^2\right)
\label{V_stephani}
\ee
In general the five functions of time in~$V$ are free. For our purposes,
however, it turns out to be more convenient to use
\be
V=a(t)+b(t)r^2-2{\mathbf{c}}(t)\cdot{\mathbf{r}},\label{Vstephani2}
\ee
as in Barnes~(1998)\nocite{barn98}, again with five free functions (we adopt
three-dimensional vector notation, so that~${\mathbf{c}}=(c_1,c_2,c_3)$, for
example). In fact, this form is slightly more general
than~(\ref{V_stephani})~-- see Barnes~(1998).

We must be able to write~$V$ in the form~(\ref{Vcondition}) for the spacetime
to admit an isotropic radiation field for all fundamental observers.
From~(\ref{V_stephani}) or~(\ref{Vstephani2}) it is clear that the functions
$X(x^i)$, and~$Y(x^i)$ in~(\ref{Vcondition}) can be at most quadratic in
the~$x^i$. Writing $X$, and~$Y$ as quadratics in~(\ref{Vcondition}) and
equating in~(\ref{Vstephani2}) all powers of~$x^i$, we obtain the following
constraint equations:
\begin{eqnarray}
                a(t) \li = \li a_1 T(t) + a_2, \nonumber \\
                b(t) \li = \li b_1 T(t) + b_2, \label{Steph_constr}\\
     {\mathbf{c}}(t) \li = \li {\mathbf{c}}_1 T(t) + {\mathbf{c}}_2, \nonumber
\end{eqnarray}
where~$T(t)$ is a free function of time and the $a_{1,2}$, $b_{1,2}$
and~${\mathbf{c}}_{1,2}$ are ten independent constants. Not all of~$a_1$, $b_1$
and~${\mathbf{c}}_1$ can be zero (in order that~$V,_t\ne 0$ in~(\ref{metric})).

Equations~(\ref{Steph_constr}), along with (\ref{metric})
and~(\ref{Vstephani2}), provide the complete set of irrotational perfect fluid
spacetimes admitting an isotropic radiation field. Not all of the possible
choices of parameters give rise to distinct spacetimes, though, and we outline
in appendix~\ref{app-trans} how coordinate transformations may be used to
eliminate many of the parameters in~(\ref{Steph_constr}), and determine when
the models can be reduced to manifestly spherically symmetric
form~(${\mathbf{c}}={\mathbf{0}}$). Given the forms for $a$, $b$,
and~$\mathbf{c}$ in~(\ref{Steph_constr}) we can equate~(\ref{V_stephani})
with~(\ref{Vstephani2}) to find the corresponding constraints on $R$, $k$,
and~${\mathbf{x}}_0$ in~(\ref{V_stephani}). This we do in
appendix~\ref{app-Rkx}.

At this point we can say that perfect fluid spacetimes admitting an isotropic
radiation field for all fundamental observers are FLRW if and only if the
acceleration of the fundamental observers is zero. This follows because the
Stephani models with zero acceleration are FLRW (see Krasi\'{n}ski~1997,
although it can easily be seen from~(\ref{acceln}): if $\dot{u}=0$ then
$V=T(t)X(x^i)$ for some functions~$T$ and~$X$, showing that $a$, $b$
and~${\mathbf{c}}$ depend on the single free function~$T$ and so must be FLRW
by the results of the next section). Thus we have proved:
\begin{thm}
\label{IPFthm}
The irrotational perfect fluid spacetimes admitting an isotropic radiation
field for the fundamental observers are Stephani models with the free functions
restricted by~(\ref{Steph_constr}) (or~(\ref{constr-app})). These spacetimes
are FLRW if and only if the acceleration of the fundamental observers is zero.
\end{thm}

It is worth noting that all of the models admitting an isotropic radiation
field are manifestly conformal to (part of) the Einstein static spacetime
(\emph{cf}.~Barrett and Clarkson~1999a) once they have been transformed so
that~${\mathbf{c}}=c{\mathbf{\hat{z}}}$ as outlined in
appendix~\ref{app-trans}. From (\ref{Vstephani2}) and~(\ref{constr-app}) we
obtain
\[
        V_{,t} = a_1T_{,t}\left( 1 + \frac{b_1}{a_1} r^2 \right),
\]
since~${\mathbf{c}}_{,t}={\mathbf{0}}$ when~${\mathbf{c}}=c{\mathbf{\hat{z}}}$
(as noted in appendix~\ref{app-trans} we can assume~$a_1\ne 0$). Changing the
time coordinate via $dt\mapsto a_1 T_{,t}F\,dt$, the metric~(\ref{metric})
becomes
\[
    ds^2  = \frac{(1+\sfrac14\Delta r^2)^2}{V(z,r,t)^2}
    \left\{ -dt^2 + \frac{1}{(1+\sfrac14\Delta r^2)^2}(dx^2+dy^2+dz^2)\right\},
\]
where~$\Delta=4b_1/a_1$. The factor in braces is the Einstein static metric.
Barrett and Clarkson~(1999a) used this conformal relationship to simplify the
study of the observational characteristics of a subset of these models.

\subsection{Symmetry and Thermodynamic Schemes.}

Having obtained the conditions~(\ref{Steph_constr}) for a Stephani model to
admit an isotropic radiation field it is possible to say immediately that all
such spacetimes possess (at least) a three-dimensional symmetry group acting on
two-dimensional orbits (just as for the general Barnes models). This follows
from the work of Barnes~(1998), who showed that the dimension of the isometry
group of any Stephani spacetime is determined by the dimension~$d$ of the
linear space spanned by the five free functions $a$, $b$, and~${\mathbf{c}}$:
\begin{enumerate}
   \item if $d=4$ or~$5$ (i.e.,~at least four of the free functions are linearly
independent), then the spacetime has no Killing vectors;
   \item if $d=3$ there is a one-dimensional isometry group;
   \item if $d=2$ there is a three-dimensional isometry group acting on
two-dimensional orbits;
   \item if $d=1$ there are six Killing vectors and the spacetime is
Robertson-Walker.
\end{enumerate}
It is clear from~(\ref{Steph_constr}) that $a$, $b$, and~$\mathbf{c}$ depend on
(at most) only two functions of time: $f_1(t)=T(t)$ and the constant
function~$f_2(t)\equiv 1$ (since~$V_{,t}\ne 0$ we must have~$T_{,t}\ne 0$, so
that these are necessarily linearly independent). Thus,~$d=2$ and the solutions
have three-dimensional isometry groups as claimed. If
$a_2=b_2={\mathbf{c}}_2=0$ in~(\ref{Steph_constr}) then~$d=1$ and the spacetime
is~FLRW (see appendix~\ref{app-trans}.

It follows further from this and the work of Bona and
Coll~(1988)\nocite{bon-col88} (see also Krasi\'nski, Quevedo and
Sussman~1997\nocite{kras-97}) that all Stephani models that admit an isotropic
radiation field for all fundamental observers also admit a strict thermodynamic
scheme (that is, entropy and temperature functions can be found that depend on
the energy density and pressure and satisfy the first law of thermodynamics as
embodied by the Gibbs equation). The converse is not true, however, since there
are Stephani spacetimes with~$d=2$ (which must admit a thermodynamic scheme)
that cannot have an isotropic radiation field (these are models for which the
second independent function~$f_2(t)$ is not restricted to be~$1$). So, we have
the following corollary to Theorem~\ref{IPFthm}:
\begin{cor}
An irrotational perfect fluid spacetime that admits an isotropic radiation
field has spherical, planar or hyperbolic symmetry and admits a strict
thermodynamic scheme.
\end{cor}

On the subject of thermodynamics, let us mention for completeness that the
thermodynamic scheme occurring most often in the literature is that of a
barotropic equation of state. It is known that the only Stephani models with a
barotropic EOS are precisely the FLRW models (Bona and Coll~1988;
Krasi\'{n}ski~1997). Thus, the only spacetimes with a barotropic EOS
admitting an isotropic radiation field are FLRW models. This also follows
(when~$\mu+p\ne 0$) from a theorem of Coley~(1991)\nocite{coley91}.

\section{QCDM Models.}
\label{QCDM}

The type~Ia supernova results of Riess \etal~(1998)\nocite{riess-98},
Schmidt \etal~(1998)\nocite{schmidt-98} and
Perlmutter \etal~(1999), which suggest that the expansion of the
universe is accelerating, have lead to an increased interest in cosmological
models in which a significant contribution to the energy density comes
from either a cosmological constant or a scalar field (quintessence
component), which is capable of driving the expansion
(Peebles and Ratra~1988\nocite{peebl-88};
Ratra and Peebles~1988\nocite{ratra-88};
Caldwell \etal~1998\nocite{caldw-98};
Zlatev \etal~1999\nocite{zlatev-99}). In QCDM models the
matter is an admixture of non-interacting cold dark matter (CDM),
i.e.,~dust, and a scalar field. The fundamental observers (galaxies)
are implicitly identified with the geodesic congruence of the
CDM. Note that there is no reason \emph{a priori} why the scalar field
gradient (which defines a natural `velocity' field) should be aligned
with the CDM congruence (although it will turn out that they are
aligned if the fundamental observers see an isotropic radiation
field).

It is interesting to ask whether the EGS theorem can be extended to this case.
We demonstrate that it can by proving the following theorem:
\begin{thm}
\label{QCDMthm}
Any solution to Einstein's equations in which the matter consists of
non-interacting radiation, expanding dust (CDM), and a scalar field (or
cosmological constant), and for which the dust sees an isotropic radiation
field, must either be an FLRW model, or have the gradient of the scalar field
orthogonal to the dust congruence.
\end{thm}
(Note that the latter possibility means that gradient of the scalar field is
spacelike, and is usually rejected as unphysical~-- although see below.)

{\raggedleft \emph{Proof:}}

We may divide this proof into two parts: first we demonstrate from Einstein's
equations in the 1+3 formalism that any energy flux component with respect to
the CDM frame must be zero if the CDM observers see isotropic radiation, then
we show that the contribution to the energy flux (with respect to the CDM
frame) from the scalar field is zero if and only if the gradient of the scalar
field is parallel (or orthogonal) to the CDM velocity~$u^a_{_{CDM}}$, so we
deduce that the velocity fields are parallel (or orthogonal). The case where
the field gradient is orthogonal to~$u^a_{_{CDM}}$ is probably unphysical, and
will be rejected. Thus, the mixture of radiation, CDM and scalar field can be
written as a single perfect fluid with geodesic fundamental
congruence~$u^a=u^a_{_{CDM}}$, and it follows from the results
of~\S\ref{IPFSolns} that the model is necessarily FLRW. Throughout this section
equation numbers of the form~(E32) will refer to the lectures of
Ellis~(1998)\nocite{ell98}.

Since the radiation is isotropic for the dust observers~$u^a$ the
energy-momentum tensor for radiation may be written in the perfect fluid
form~(\ref{perfectfluid}) with~$p=\sfrac13\mu$ and the total energy-momentum
tensor is:
\be
T_{ab}=\underbrace{\mu u_au_b+\sfrac{1}{3}\mu
h_{ab}}_{\textrm{Radiation}}+\underbrace{\vphantom{\sfrac13}\rho
u_au_b}_{\textrm{CDM}} +\underbrace{\phi_{,a}\phi_{,b}
-g_{ab}\left(\sfrac{1}{2}\phi_{,c}\phi^{,c}+\Phi(\phi)\right)}_{\textrm{Scalar
Field}},
\label{T-QCDM}
\ee
where $\Phi(\phi)$ is the scalar field potential (often assumed to be zero, in
which case the scalar field can be interpreted as a stiff perfect fluid). Note
that the cosmological constant case can be included by setting
$\phi=\Lambda=$constant,~$\Phi(\phi)=\phi$.

\begin{enumerate}

\item
The fundamental congruence  $u^a$ is geodesic ($\udot^a=0$) because the CDM
component does not interact with the other matter. This, in fact, implies that
the rotation of~$u^a$ must also vanish: from the momentum conservation equation
for the radiation, we can write
\[
\udot_a=-\frac14\sdel_a\ln\mu=0
\]
(where $\sdel_a$ denotes the spatially projected gradient), so that (using
(E27))
\[
0=\sdel_{[a}\udot_{b]}=\frac14\sdel_{[b}\sdel_{a]}\ln\mu=
\frac14\omega_{ab}\frac{\dot\mu}{\mu}=\frac13\omega_{ba}\theta,
\]
and we see that $\omega_{ab}=0$ when~$\theta\ne 0$.

When $\dot{u}_a$ and~$\omega_{ab}$ are zero, (\ref{Qcondition}) becomes
\[
\del_{[a}(\theta u_{b]})= u_{[b}\del_{a]}\theta = 0.
\]
This implies (since $\del_a\theta=\sdel_a\theta-\dot{\theta}u_a$) that
\be
\sdel_a\theta=0.
\label{gradexp}
\ee
(i.e.,~the expansion is homogeneous). From the constraint equation relating the
divergence of the shear to other kinematical quantities (E32) we see that any
energy flux component with respect to the CDM velocity field must vanish:
\be
q_a=\frac23\sdel_a\theta=0.
\label{qa}
\ee
This is the key step in the proof.

\item
Decomposing ~(\ref{T-QCDM}) with respect to $u^a$ we find that the relative
energy flux component is
\be
0=q_a=-h_a^{~b}u^cT_{bc}=-\dot\phi\sdel_a\phi.
\ee
So $q_a=0$ if $\dot{\phi}=0$ (the scalar field gradient is orthogonal to~$u^a$,
and therefore spacelike), or if~$\sdel_a\phi=0$ (the scalar field gradient is
parallel to~$u^a$). We take the latter case to be most important since the
gradient of a scalar field is usually assumed to be timelike.

Since $\del_a\phi=\sdel_a\phi-u_a\dot\phi=-u_a\dot\phi$ it is possible to
write~(\ref{T-QCDM}) in the form of a single perfect fluid~(\ref{perfectfluid})
with geodesic, shear-free, rotation-free velocity field; it is thus an
FLRW model by~\S\ref{IPFSolns}. {\hfill $\Box$}

\end{enumerate}

It is easy to see from the above proof that the fact that the fundamental
observers correspond to dust-like matter was not used, only that they followed
geodesics. Hence the above result applies for more general perfect fluids in
place of the CDM component, as long as the fundamental congruence is geodesic.

The idea of a spacelike scalar field gradient seems physically unappealing.
However, such a field can (depending on the potential~$\Phi$) satisfy the weak,
strong and dominant energy conditions. The strong energy condition will be
satisfied if and only if~$\Phi(\phi)\le 0$ everywhere, whereas as the weak and
dominant energy conditions will be satisfied if~$\Phi(\phi)\ge 0$, although not
only so. Thus, a massless scalar field (stiff perfect fluid) with spacelike
gradient satisfies all energy conditions. It should be borne in mind, though,
that scalar fields arising in cosmological contexts often fail to satisfy the
energy conditions. This case may deserve further consideration. As can easily
be seen, the scalar field component gives rise to anisotropic stresses in the
energy-momentum tensor, so such spacetimes are not FLRW.

\section{Conclusions.}
\label{concs}

We have proved that the irrotational perfect fluid spacetimes admitting an
isotropic radiation field are Stephani models restricted
by~(\ref{Steph_constr}) (see also equations~(\ref{constr-app})), and are FLRW
if and only if the acceleration~$\dot{u}^a$ of the fundamental congruence is
zero (Theorem~\ref{IPFthm}). It follows from the fact that the
constraints~(\ref{Steph_constr}) depend on only two independent functions of
time that all of the acceptable models possess three-dimensional symmetry
groups acting on two-dimensional orbits (i.e.,~have spherical, planar, or
hyperbolic symmetry) and therefore possess a thermodynamic interpretation. We
have also shown that spacetimes containing a mixture of radiation, dust and
scalar field (QCDM models) for which the dust observers see the radiation as
isotropic must always be homogeneous and isotropic (Theorem~\ref{QCDMthm})
unless the scalar field gradient is spacelike and orthogonal to the CDM
congruence~-- a possibility we reject as unphysical. This result also relies on
the geodesic nature of the fundamental congruence.

Crucial, therefore, to the proof of homogeneity and the verification of the
cosmological principle is the non-acceleration of the fundamental observers.
Despite the intuitive appeal of cosmological models in which the fundamental
observers are associated with a dust-like matter component, it is unacceptable
to simply assume that we are geodesic observers, especially when such an
assumption is, in principle, testable. Acceleration leaves a characteristic
dipole signature in the redshifts of nearby galaxies that may be detectable
using galaxy surveys. The physical principle underlying this effect is easy to
see. For a set of uniformly accelerated observers (`galaxies') in Minkowski
space each observer will see other galaxies redshifted or blueshifted in a
dipole pattern, with the blueshifted galaxies lying in the direction of the
acceleration, because during the light-travel time between galaxy and observer
the observer's velocity has increased relative the velocity at emission, so
that the galaxies the observer is travelling towards are blueshifted, and those
it is travelling away from redshifted. It also follows from this that the
magnitude of the dipole increases with distance, simply because the
light-travel time from more distant galaxies is larger. In a cosmological
context this acceleration dipole must be added to other terms contributing to
the redshift of nearby objects, in particular the expansion. The method of
Kristian and Sachs (1966)\nocite{kris-sac66} and MacCallum and Ellis
(1970)\nocite{Mac-Ellis-II} gives, for any cosmological model with fundamental
congruence~$u^a$, the lowest-order term in the redshift~$z$ as a function of
distance~$r$ and direction~$e^a$ (a spacelike unit vector orthogonal to~$u^a$,
denoting the direction of observation):
\be
z=
\left. H_0r\left(1-\frac{\udot_ae^a}{H_0}+\frac{\sigma_{ab}e^ae^b}{H_0}\right)
\right|_0+{\cal{O}}(r^2),
\label{H0}
\ee
where $H_0=\sfrac13\theta_0$ is Hubble's constant and the last term in brackets
indicates the quadrupole introduced by the presence of shear. In~(\ref{H0}) $r$
can be any cosmological distance measure (area distance, for example) because
for small~$r$ all such measures agree to first order. Note that just as the
monopole (expansion) term increases linearly with distance according to the
Hubble law, so does the acceleration dipole. This is important, because it
allows the acceleration dipole to be distinguished from any dipole resulting
from the peculiar velocity of our galaxy with respect to the cosmological
average rest frame (usually identified with the CMB frame). Equation~(\ref{H0})
applies in this rest frame, and any peculiar motion results in a doppler shift
for each galaxy, which introduces an additional dipole component into the
galaxy redshifts. This dipole is just a constant depending only on the peculiar
velocity of our galaxy. A boost to the `correct' rest frame can eliminate this
constant component, but cannot remove the acceleration dipole because it is
distance dependent. It is important to note in this context that the
acceleration referred to here is not the same as the `acceleration dipole'
resulting from the gravitational attraction by the Great Attractor overdensity,
which is often calculated using galaxy surveys
(see Schmoldt \etal~1999\nocite{sch-99}).

Galaxy surveys are often used to measure a possible bulk flow of the local
universe, that is, the difference, if any, between the rest frame of the local
universe and the CMB frame, which in standard cosmological models should be the
same (see Willick 1998\nocite{willick}). A simple extension of these techniques
(Clarkson, Rauzy and Barrett, in preparation) permits the acceleration to be
constrained by observations. However, preliminary results suggest that the
constraints on~$\dot{u}^a$ are quite weak: it appears not to be possible to
conclude definitively that we are geodesic observers. The accuracy
of~$\dot{u}^a$ determinations is limited both by uncertainties in the distance
estimates to galaxies as well as the peculiar velocities of galaxies.

Even if the acceleration was measured to be zero, it is still necessary to show
that there are no anisotropic stresses (Ferrando \etal~1992) before the
cosmological principle can be verified. It follows from equation~(31) of
Ellis~(1998) that this is equivalent to determining that the electric part of
the Weyl tensor is zero. It is not clear how this may be achieved using
observations.

Of course, the Copernican principle (which is a vital element of EGS-type
theorems, allowing the high isotropy of the CMB here to be assumed for other
points in the universe) remains a purely philosophical assumption. It has been
suggested by a number of authors (see Goodman~1995, for example) that
the Sunyaev-Zeldovich effect might be used to place constraints on the
anisotropy of the CMB at distant positions, but it is not obvious that
such observations will provide a definitive verification of the Copernican
principle. Nevertheless, the arguments in favour
of the Copernican principle are quite powerful, and it is a much weaker
assumption than the cosmological principle. Note that if the acceleration
\emph{is} measured to be zero here, the Copernican principle must also be
applied to give geodesic observers everywhere for the results of this paper
(and the other EGS papers) to hold.

Finally, one might expect that the `almost' version of Theorem~\ref{IPFthm}
would lead to spacetimes that are almost the Stephani models
of~(\ref{Steph_constr}). However, when the assumption of geodesic observers is
relaxed it is no longer possible to constrain the rotation to be small, and the
class of perfect fluid spacetimes with an almost isotropic CMB may include
examples with distinctly non-zero rotation, unless other constraints are
brought to bear. Actually, it is possible that \emph{all} spacetimes admitting
an IRF are irrotational (see Coley~1991,~\S2.2), although this has not been shown
definitively. It would be interesting to determine the class of all
perfect fluid spacetimes admitting an isotropic radiation field, and if it turns
out that they are indeed all irrotational then the Stephani spacetimes defined
by~(\ref{Steph_constr}), (\ref{metric}) and~(\ref{Vstephani2}) are indeed
the complete set.

\section*{Acknowledgements}

We would like to thank Roy Maartens for a great deal of input and
encouragement, Bruce Bassett for being very helpful, and for looking out for
the `little fish,' Stephane Rauzy for useful discussions and Alan Coley for
helpful suggestions. CAC was funded by a PPARC Studentship.


\newpage

\appendix

\section{Coordinate transformations and the Stephani Spacetimes.}
\label{app-trans}

On the face of it the Stephani models admitting an isotropic radiation field
in~(\ref{Steph_constr}) depend on one free function and ten free parameters.
However, it is possible to use coordinate transformations on the spacetime to
eliminate many of these parameters, resulting in a considerable simplification.
As is shown in Barnes~(1998), conformal transformations of the coordinates on
the hypersurfaces of constant time preserve the form of the metric but change
the free functions $a$, $b$, and~$\mathbf{c}$. These transformations can be
thought of as acting on the five-dimensional space spanned by $a$, $b$
and~${\mathbf{c}}$ and constitute the Lorentz group in five
dimensions,~$SO(4,1)$: they leave~$-ab + |{\mathbf{c}}|^2$ invariant ($a$
and~$b$ are `null coordinates'). It will be convenient here to let
$a=\alpha+\beta$ and~$b=\alpha-\beta$ (so that the transformations preserve
$-\alpha^2 + \beta^2 + |{\mathbf{c}}|^2$), and to adopt five-vector
notation:~$q^\mu=(\alpha,\beta,{\mathbf{c}})$. We will use the terms `rotation'
and `boost' to refer to the transformations on~$q$, and will call~$q$ timelike,
spacelike or null if $-\alpha^2 + \beta^2 + |{\mathbf{c}}|^2$ is negative,
positive or zero, as usual. Then it is easy to visualise the transformations on
the free functions by imagining the `mass hyperboliods' of representations of
the Lorentz group in the usual way: a timelike vector can always be boosted so
that it has the form~$(\alpha,0,{\mathbf{0}})$, whereas a spacelike vector can
be boosted and rotated into~$(0,\beta,{\mathbf{0}})$, for example.

In addition to the Lorentz transformations we also have the freedom to change
basis in the function space spanned by the free functions. For the spacetimes
of interest here, described by~(\ref{Steph_constr}), it is desireable to
preserve~$f_2(t)=1$, so that the basis change is
\be
                  T \mapsto \gamma T + \delta.
\label{Ttrans}
\ee

In five-vector notation the equations~(\ref{Steph_constr}) become
\be
    q^\mu(t) \equiv (\alpha,\beta,{\mathbf{c}}) = q_1^\mu T(t) + q_2^\mu
\label{Steph_conq}
\ee
(with $q_1$ and~$q_2$ constant vectors), and the goal is to reduce as many of
the components of $q_1^\mu$ and~$q_2^\mu$ to zero as possible using Lorentz
transformations in the five-dimensional space containing $q_1$ and~$q_2$, and
the basis change~(\ref{Ttrans}). Note that the FLRW ($d=1$) subcase
of~(\ref{Steph_constr}) is characterised by the linear dependence of $q_1$
and~$q_2$.

It is easy to see that~${\mathbf{c}}$ (which breaks the spherically symmetry of
the metric~(\ref{metric})) may always be reduced to the
form~${\mathbf{c}}=c{\mathbf{\hat{z}}}$ (${\mathbf{\hat{z}}}=(0,0,1)$),
with~$c$ a constant: perform a spatial 4-rotation to reduce~$q_1^\mu$
to~$q_1^\mu=(\alpha_1,\beta_1,{\mathbf{0}})$, then a spatial rotation amongst
the ${\mathbf{c}}$-components (which obviously leaves~$q_1$ unaffected) to
give~$q_2^\mu=(\alpha_2,\beta_2,c{\mathbf{\hat{z}}})$.

It is possible in general to make further simplifications, but precisely how
$q_1$ and~$q_2$ are simplified depends on whether they are spacelike, timelike
or null. For example, if either $q_1$ or~$q_2$ is timelike (or may be made
timelike by a transformation~(\ref{Ttrans})) it is possible to reduce the model
to manifestly spherically symmetric form ($c=0$): boost so that the timelike
vector, say~$q_1$, becomes~$q_1=(\alpha_1,0,{\mathbf{0}})$ and rotate spatially
so that the ${\mathbf{c}}$-components of the other vector are also
zero,~$q_2=(\alpha_2,\beta_2,{\mathbf{0}})$ (we could then use~(\ref{Ttrans})
to eliminate more of these constants).

To summarise, we have demonstrated that it is always possible to reduce the
Stephani models of~(\ref{Steph_constr}) to the form
\begin{eqnarray}
                a(t) \li = \li a_1 T(t) + a_2, \nonumber \\
                b(t) \li = \li b_1 T(t) + b_2, \label{constr-app}\\
     {\mathbf{c}}(t) \li = \li c{\mathbf{\hat{z}}}, \nonumber
\end{eqnarray}
(where we have transformed back from $\alpha$ and~$\beta$ to $a$ and~$b$), and
when either of the $q_1$ or~$q_2$ is (or may be made) timelike we can
set~$c=0$.

Finally, note that we may always assume that~$a_1\ne 0$ in~(\ref{constr-app}),
because if~$a_1=0$ then~$b_1\ne 0$, otherwise~$V_{,t}=0$, and it is possible to
perform a coordinate inversion ${\mathbf{x}}\mapsto {\mathbf{x}}/r^2$ that
interchanges $a$ and~$b$). This does not exhaust the possibilities for
simplification: we could, for example, use~(\ref{Ttrans}) to set~$a_2=0$.

\section{The Constraints on $R$, $k$ and~${\mathbf{x}}_0$.}
\label{app-Rkx}

To find the constraints on $R$, $k$ and~${\mathbf{x}}_0$ in~(\ref{V_stephani})
corresponding to~(\ref{Steph_constr}) first equate powers of~$x^i$ in
(\ref{V_stephani}) and~(\ref{Vstephani2}) to obtain
\begin{eqnarray}
   a(t)\li=\li \frac{1}{R}+\frac{k}{4R}|{\mathbf{x}}_0|^2, \label{A}\\
   b(t)\li=\li \frac{k}{4R}, \label{B}\\
   {\mathbf{c}}(t)\li=\li \frac{k}{4R}{\mathbf{x}}_0. \label{C}
\end{eqnarray}
Solving these equations for $R$, $k$ and~${\mathbf{x}}_0$ gives
\begin{eqnarray}
   R(t) \li=\li \frac{b}{ab-|{\mathbf{c}}|^2}, \label{R}\\
   \frac14 k(t) \li=\li \frac{b^2}{ab-|{\mathbf{c}}|^2}, \label{k}\\
   {\mathbf{x}}_0 \li=\li \frac{\mathbf{c}}{b}, \label{x}
\end{eqnarray}
which are valid whenever ${ab-|{\mathbf{c}}|^2}\ne 0$ (otherwise $V$ cannot be
written in the form~(\ref{V_stephani})).

To impose the constraints~(\ref{Steph_constr}) perform the transformations of
appendix~\ref{app-trans} so that ${\mathbf{c}}=c{\mathbf{\hat{z}}}$ as
in~\ref{constr-app}) and~$a_1\ne 0$. Then $|{\mathbf{c}}|^2 = c^2$, and $b$
and~$a$ are related by
\be
          b = \frac{b_1}{a_1}a + \left(b_2-\frac{b_1a_2}{a_1}\right)
            \equiv \gamma a+\delta,
\label{bofa}
\ee
where $\gamma=b_1/a_1$ and~$\delta=b_2-b_1a_2/a_1$ are constants. Using this in
(\ref{R}) and~(\ref{k}) leads, after some rearrangement, to a quadratic
relationship between $k$ and~$R$:
\be
    \left(\frac{k}{4}\right)^2 - (\gamma + \delta R)\left(\frac{k}{4}\right)
              - \gamma c^2 R^2 = 0.
\label{kReqn}
\ee
In addition to this constraint relating $k$ and~$R$ we can trivially rewrite
(\ref{C}) or~(\ref{x}) as
\be
          {\mathbf{x}}_0 = \frac{4R}{k} c{\mathbf{\hat{z}}}.
\label{ckR}
\ee

Equations (\ref{kReqn}) and~(\ref{ckR}) are the constraint equations on $R$,
$k$ and~${\mathbf{x}}_0$ corresponding to the equations~(\ref{Steph_constr}),
or rather~(\ref{constr-app}). If desired (\ref{kReqn}) can be solved to obtain
\[
         \frac{k}{4} = \frac12 \left[\gamma+\delta R
                       \pm \sqrt{(\gamma+\delta R)^2 + 4\gamma c^2R^2}\right].
\]

From~(\ref{kReqn}) it is clear that the spherically symmetric Stephani
spacetimes admitting an isotropic radiation field, for which~$c=0$, satisfy
\[
   \frac{k}{4}\left(\frac{k}{4} - (\gamma + \delta R)\right) = 0,
\]
which has the solutions $k=0$ and $R(t)$~free (flat Friedmann), or $k$ linearly
related to~$R$,
\[
         \frac14 k(t)=\gamma +\delta R(t),
\]
with $R(t)$ again free (when~$\delta=0$ these become Friedmann models with
curvature~$k=4\gamma$). The Stephani models studied by Barrett and
Clarkson~(1999a,b) are members of this class (with $R$ depending quadratically
on~$t$), which explains the isotropy of the microwave background found for
those spacetimes.

\end{document}